\documentclass[twocolumn,showpacs,preprintnumbers,amsmath,amssymb]{revtex4}

\usepackage{graphicx}
\usepackage{dcolumn}
\usepackage{bm,textcomp}
\newcommand*{\micro}{\ensuremath{\text{\textmu}}}

\begin{document}


\title{Simulating the Quantum Magnet}

\author{Axel Friedenauer}
\author{Hector Schmitz}
\author{Jan Gl\"uckert}
\author{Diego Porras}
\author{Tobias Sch\"atz}
 \email{tobias.schaetz@mpq.mpg.de}
\affiliation{
Max-Planck-Institut f\"ur Quantenoptik, Hans-Kopfermann-Stra\ss{}e 1, 85748 Garching, Germany\\
}%

\date{\today}

\begin{abstract}
To gain deeper insight into the dynamics of complex quantum systems we need a quantum leap in computer simulations. We can not translate quantum behaviour arising with superposition states or entanglement efficiently into the classical language of conventional computers. The final solution to this problem is a universal quantum computer \cite{1}, suggested in 1982 and envisioned to become functional within the next decade(s); a shortcut was proposed via simulating the quantum behaviour of interest in a different quantum system, where all parameters and interactions can be controlled and the outcome detected sufficiently well. 

Here we study the feasibility of a quantum simulator based on trapped ions \cite{2}. We experimentally simulate the adiabatic evolution of the smallest non-trivial spin system from the paramagnetic into the (anti-)ferromagnetic order with a quantum magnetisation for two spins of 98\%, controlling and manipulating all relevant parameters of the Hamiltonian independently via electromagnetic fields. We prove that the observed transition is not driven by thermal fluctuations, but of quantum mechanical origin, the source of quantum fluctuations in quantum phase transitions \cite{3}. We observe a final superposition state of the two degenerate spin configurations for the ferromagnetic ($\left| \uparrow \uparrow \right\rangle + \left| \downarrow \downarrow \right\rangle$) and the anti-ferromagnetic ($\left| \uparrow \downarrow \right\rangle + \left| \downarrow \uparrow \right\rangle$) order, respectively. These correspond to deterministically entangled states achieved with a fidelity up to 88\%.    

Our work demonstrates a building block for simulating quantum spin-Hamiltonians with trapped ions. The method has potential for scaling to a higher number of coupled spins \cite{2}. 
\end{abstract}

\maketitle

\section{Introduction}		

It is not possible to efficiently describe the time evolution of quantum systems on a classical device, like a conventional computer, since their memory requirements grow exponentially with their size. For example, a classical memory needs to hold $2^{50}$ numbers to store arbitrary quantum states of 50 \mbox{spin-1/2} particles. To be able to calculate its evolution demands to derive a matrix of \mbox{$(2^{50})^2=2^{100}$} elements, already exceeding by far the capacity of state of the art computers. Each doubling of computational power permits only one additional \mbox{spin-1/2} particle to be simulated. To allow for deeper insight into quantum dynamics we need a ``quantum leap'' in simulation efficiency.
\begin{figure}
\includegraphics[width=8cm]{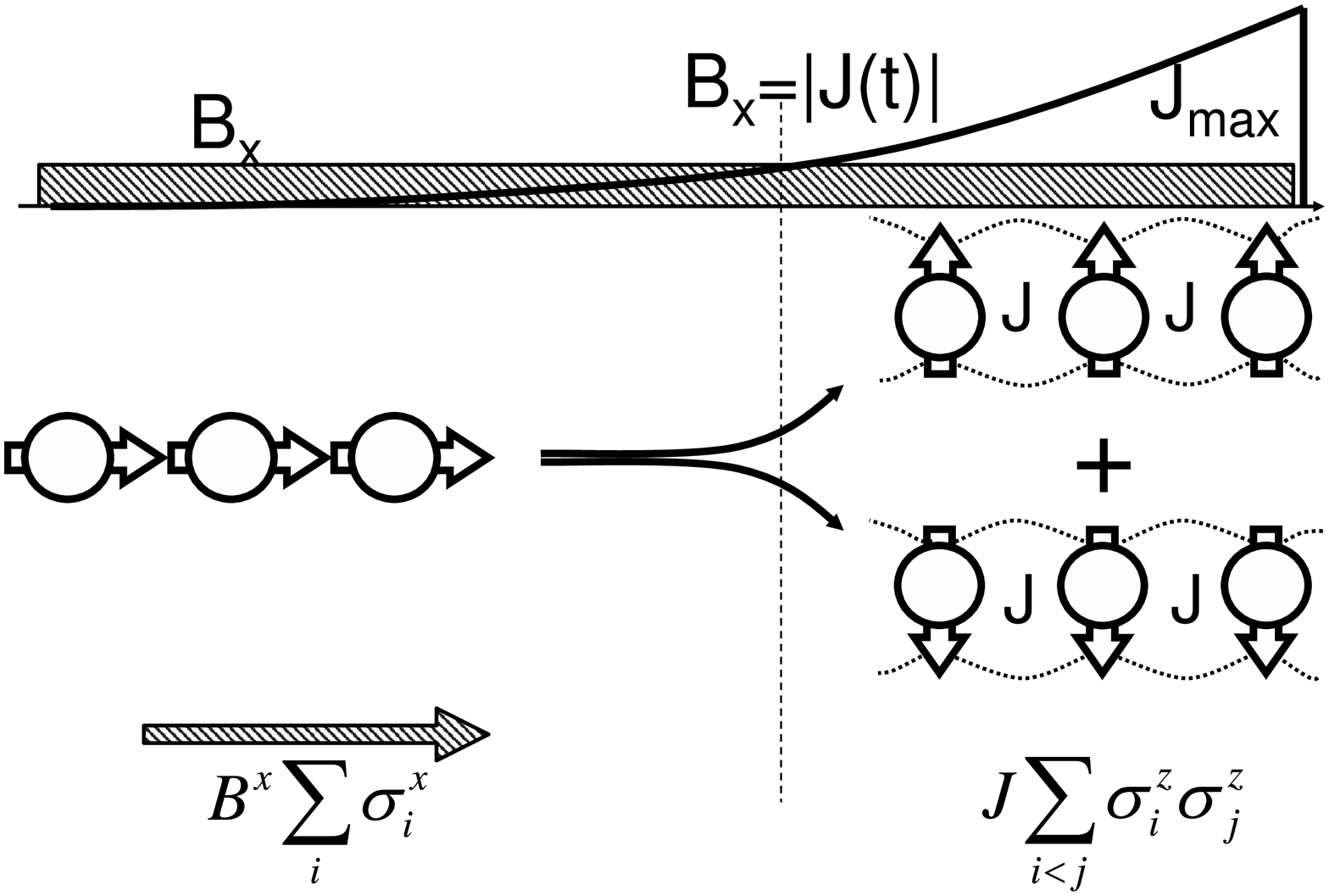}
\caption{Phase transition of a quantum magnet:
Each ion can simulate a magnetic spin, analogue to an elementary magnet. We initialise the spins in the paramagnetic state $\left| \rightarrow \rightarrow \ldots \rightarrow \right\rangle$, the ground state of the Hamiltonian \mbox{$H_\text{B}=B_x(\sigma_1^x +\sigma_2^x+ \ldots + \sigma_N^x)$}. This is equivalent to aligning the spins parallel to the simulated magnetic field.
Adding an effective spin-spin interaction $J(t)$ (at constant $B_x$) and increasing it adiabatically to $|J_\text{max}|\gg B_x$ , we expect the system to undergo a quantum phase transition into a ferromagnet, the new ground state of the system ($J$ is symbolized as little chains, trying to keep neighbouring spins aligned). Ideally, the two possible ferromagnetic orders $\left| \uparrow \uparrow \ldots \uparrow \right\rangle$ and $\left| \downarrow \downarrow \ldots \downarrow \right\rangle$ are degenerate ground states. The spin system should evolve into the superposition state $\left| \uparrow \uparrow \ldots \uparrow \right\rangle + \left| \downarrow \downarrow \ldots \downarrow \right\rangle$, a maximally entangled ``Schr\"odinger Cat'' state/magnet.} \label{anschaulich} 
\end{figure}

As proposed by Richard Feynman \cite{1}, a universal quantum computer would accomplish this step.  A huge variety of possible systems are under investigation, a very promising one being trapped ions \cite{6,7} acting as quantum bits (qubits). After addressing the established criteria summarized by DiVincenzo \cite{8} on up to 8 ions \cite{9,10} with operational fidelities exceeding 99\% \cite{9,10,11}, there seems to be no fundamental reason why such a device would not be realisable.

An analogue quantum computer, much closer to the original proposal by Feynman, might allow for a shortcut towards quantum simulations. We want to simulate a system by a different one being described by a Hamiltonian that contains all important features of the original system. The simulator needs to be  controlled, manipulated and measured in a sufficiently precise manner and has to be rich enough to address interesting questions about the original system. For large coupled spin systems optical lattices might be advantageous \cite{13}, while smaller spin systems and degenerate quantum gases might be simulated by trapped ions \cite{2,14}. Instead of implementing a Hamiltonian with a universal set of gates, direct simulation of the Hamiltonian typically consists of one (adiabatic) evolution of the initial state into the corresponding final state of interest.
 
Here, in a proof-of-principle experiment, we simulate the adiabatic transition from a quantum para- to a quantum (anti-)ferromagnet and illustrate the advantages of the adiabatic quantum simulation (see FIG.~\ref{anschaulich}). We demonstrate the individual access, via rf- and laser fields, to all relevant parameters in the underlying Hamiltonian, representing one out of a large spectrum of quantum spin-Hamiltonians.

\section{Adiabatic Quantum Simulation}

\begin{figure}
\includegraphics[width=8cm]{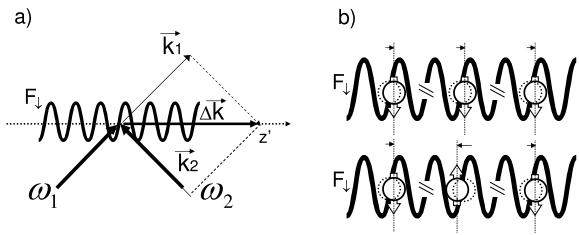}
\caption{\label{dipole} Simulating the quantum magnet:
a) Two perpendicular polarized laser beams of frequency $\omega_1$ and $\omega_2$ induce a state dependent optical dipole force $F_\downarrow = -3/2 \, F_\uparrow$ along the trap axis $a$ by the AC-Stark shift (here, only $F_\uparrow$ is depicted). b) For a standing wave $\omega_1=\omega_2$, the force conditionally changes the distance between neighbouring spins simulating a spin-spin interaction \cite{2} ($F_\uparrow$ ($F_\downarrow$) symbolised by the arrow to the right (left)). Only if all spins are aligned (top), the total Coulomb energy of the spin system is not increased, defining ferromagnetic order, the quantum magnet, to be the ground state. For $\omega_1\neq \omega_2$, the sinusoidal force pattern can be seen as a wave moving along the trap axis $a$ pushing or pulling the ions repeatedly at a frequency $\omega_1-\omega_2$. We chose $\omega_1-\omega_2$ close to the resonance frequency of the ions oscillating out of phase (stretch mode). The energies of different spin states now depend on the coupling of the spin state to the stretch mode. Energy can be coupled efficiently into the state with different spin orientations (e.g. bottom), defining the not affected upper case as a ground state \cite{18}. The interpretation in terms of an effective spin-spin interaction is further described in methods.}
\end{figure}

The adiabatic quantum simulation of generic spin-Hamiltonians proposed by Porras and Cirac \cite{2} can be illustrated considering a string of charged spin-1/2 particles confined in a common harmonic potential. Two electronic states of each ion simulate the two-level system of a spin-1/2 magnetic moment, $\left| \uparrow \right\rangle$ and $\left| \downarrow \right\rangle$. Note that the inter-ion distance of several $\micro$m renders any direct spin-spin coupling negligible. The quantum Ising Hamiltonian, \begin{equation} H_\text{Ising}=H_\text{B} + H_\text{J} = B_x \sum_i \sigma_i^x + \sum_{i<j} J_{ij} \sigma_i^z \sigma_j^z \ , \end{equation} consists of two terms. The first denotes the interaction of each individual spin, represented by the Pauli operator $\sigma_i^k$ ($k$ can be $x$, $y$, or $z$), with a uniform magnetic field $B_x$ pointing into direction $x$. The second term stands for the spin-spin interaction which tries to align the spins ($\sigma_i^z$) parallel or anti-parallel along the $z$-axis dependent on the sign of the interaction amplitude $J_{ij}$. To simulate the first one they couple the eigenstates of $\sigma_i^z$, $\left| \uparrow \right\rangle$ and $\left| \downarrow \right\rangle$, via an electromagnetic field. The latter is simulated by a state-dependent forcing \cite{16,4}, further explained with the help of FIG.~\ref{dipole}. 

To understand the experiment discussed below, we can consider two extreme scenarios and interactions between nearest neighbours only, $J_{i,i+1}=J$. For the case of $J=0$ and $B_x > 0$, the ground state of the spin-system has all spins aligned with $B_x$ along the $x$-axis. This corresponds to the paramagnetically ordered state $\left| \rightarrow \rightarrow \ldots \rightarrow \right\rangle$, the eigenstate of the Hamiltonian $B_x \sum_i \sigma_i^x$ with the lowest energy.
 
For the opposite case of $B_x\! =\!0$ and $J\!<\!0 \ (J\!>\!0)$,\linebreak the system has an infinite number of degenerate ground states, defined by any superposition of the lowest energy eigenstates of $\sum_{i<j} J_{ij} \sigma_i^z \sigma_j^z$, namely $\left| \uparrow \uparrow \ldots \uparrow \right\rangle$ and $\left| \downarrow \downarrow \ldots \downarrow \right\rangle$ which represents ferromagnetic order (or $\left| \uparrow \downarrow \uparrow \ldots \uparrow \downarrow \right\rangle$ and $\left| \downarrow \uparrow \downarrow \ldots \downarrow \uparrow \right\rangle$, the anti-ferromagnetic order, respectively). Initialising the spin system in an eigenstate in the $\sigma^x$-basis, starting with \mbox{$J(t\!=\!0)=0 $} and $B_x > 0$ and adiabatically increasing $\left| J(t) \right|$ to \mbox{$\left| J_\text{max} \right| \gg B_x$} should evolve the system from the paramagnetic arbitrarily close into the (anti-)ferromagnetic order, as depicted in FIG.~\ref{anschaulich}. A quantum phase transition is supposed to occur at $B_x=\left| J \right|$ in the thermodynamic limit of an infinite amount of spins \cite{3,21}.

\section{Experimental implementation}
We implemented the experiment as described in the following. We confine two $^{25}\text{Mg}^+$ ions in a linear Paul trap \cite{17} and laser-cool them to the Coulomb-crystalline phase where the ions align along the trap axis $a$. The motion of the ions along $a$ can be described in the basis of normal modes: the oscillation-in-phase-mode (com) and the oscillation-out-of-phase-mode (stretch). The related oscillation frequencies amount to $\omega_\text{com}=2\pi \times2.1$ MHz and $ \omega_\text{stretch}=2\pi \times3.7$ MHz, respectively.

In our implementation we define the hyperfine ground states $\left| \downarrow \right\rangle \equiv \left| F=3 ,\, m_f=3 \right\rangle$ and $\left| \uparrow \right\rangle \equiv \left| F=2 \, , m_f=2 \right\rangle$ in the $2\, \text{S}_{1/2}$ levels separated by $\omega \cong 2 \pi \times 1.7~\text{GHz}.$  An external magnetic field $B$ of $5.5~\text{G}$ orients the magnetisation axis for the projection $\hbar m_f$ of each ion's angular momentum $F$.  In this field adjacent Zeeman sublevels of the $F=3$ and $F=2$ manifolds split by $2.7$ MHz per level.

We coherently couple the states $\left| \downarrow \right\rangle$ and the $\left| \uparrow \right\rangle$ with a resonant radio-frequency field at $\omega_0$ to implement single spin rotations \cite{15,18}, \begin{equation} \begin{aligned} R(\Theta,\phi) &= \cos(\Theta/2) \, I -i \sin(\Theta/2) \cos(\phi) \, \sigma^x \\ &-i \sin(\Theta/2) \sin (\phi) \, \sigma^y \, ,\end{aligned} \end{equation} where $I$ is the identity operator, $\sigma^x$ and $\sigma^y$ denote the Pauli spin matrices acting on $\left| \downarrow \right\rangle$ and $\left| \uparrow \right\rangle$, $\Theta/2=B_x t$ is proportional to the duration $t$ of the rotation and $\phi$ is the phase of the rf-oscillation, defining the axis of rotation in the $x$-$y$-plane of the Bloch sphere.

We provide the effective spin-spin interaction by a state dependent optical dipole force \cite{16,18,22}. The relative amplitudes $F_\downarrow = -3/2\, F_\uparrow$ are due to AC-Stark shifts induced by two laser beams at wavelength $\lambda$ of 280 nm, depicted in FIG.~2a, perpendicular in direction and polarisation with their effective wave-vector difference pointing along the trap axis $a$. They are detuned 80 GHz blue of the $^2\text{P}_{3/2}$ excited state, with intensities allowing $J/\hbar$ above $2 \pi \times 22.1$ kHz.  We use a walking wave force-pattern by detuning the two laser wavelengths by $2\pi \times 3.45\ \text{MHz}= \omega_\text{stretch} +\delta$ with $\delta=-2\pi \times 250\ \text{kHz}$. This choice avoids several technical problems of the original proposal \cite{2} (see methods), while at the same time, resonantly enhancing the effective spin-spin interaction by a factor of $ \left| \omega_\text{stretch} / \delta  \right | =14.8$ compared to the standing wave case \cite{25}. 

After laser cooling we initialise the quantum simulator by optical pumping \cite{19} to the state $\left| \downarrow \right\rangle \! \left| \downarrow \right\rangle \left| n\cong0 \right\rangle$. We rotate both spins in a superposition state via a $R(\pi/2,-\pi/2)$-pulse (see Eq.\,2) on the rf-transition to initialise the state $\Psi_\text i =\left| \rightarrow \right\rangle\! \left| \rightarrow \right\rangle \! \left| n\cong 0 \right\rangle$. Note that this paramagnetic state $\left| \rightarrow \right\rangle \! \left| \rightarrow \right\rangle\equiv (\left| \uparrow \right\rangle+\left| \downarrow \right\rangle)(\left| \uparrow \right\rangle+\left| \downarrow \right\rangle)= \left| \uparrow \uparrow \right\rangle+\left| \uparrow \downarrow \right\rangle+\left| \downarrow \uparrow \right\rangle+\left| \downarrow \downarrow \right\rangle$ has a 25\% probability to be projected into either $\left| \uparrow \uparrow \right\rangle$ or $\left| \downarrow \downarrow \right\rangle$ (normalisation factors are suppressed throughout).

After the adiabatic evolution described below, we project the final spin state into our $\sigma^z$-measurement basis by a laser beam tuned resonantly to the \mbox{$\left| \downarrow \right\rangle \leftrightarrow {^2\text{P}}_{3/2} \left| F=4 , \, m_f=4 \right\rangle$} cycling transition \cite{18}. An ion in state $\left| \downarrow \right\rangle$ fluoresces brightly, leading to the detection of on average 40 photons during a $160~\micro \text s$ detection period with our photo multiplier tube. In contrast, an ion in state $\left| \uparrow \right\rangle$ remains close to dark (on average 6 photons). We repeat each experiment for the same set of parameters $10^4$ times and derive the probabilities $P_{\downarrow \downarrow},P_{\uparrow \uparrow}$ and $P_{\downarrow \uparrow}$ for the final state being projected into state $\left| \downarrow \downarrow \right\rangle,\ \left| \uparrow \uparrow \right\rangle$ and $\left| \downarrow \uparrow \right\rangle$ or $\left| \uparrow \downarrow \right\rangle$, respectively (and further described in methods).

We simulate the effective magnetic field by continuously applying a radio-frequency field with phase $\phi=0$ and an amplitude such that it corresponds to a single qubit rotation $R(\Theta,0)$ with full rotation period $\Theta=2\pi$ in $118~\micro \text s$ and deduce $B_x=2\pi \times 4.24~\text{kHz}$. Precise control of the phase $\phi$ of the rf-oscillator relative to the initialisation pulse allows to align $B_x$ parallel to the spins along the $x$-axis in the equatorial plane of the Bloch sphere, ensuring that $\left| \Psi_\text i \right\rangle$ is an eigenstate of this effective magnetic field.

At the same time, we switch on the effective spin-spin interaction $J(t)\ (t \in [0 ; T] )$ and increase its amplitude adiabatically up to $J(T)$. At time $T$, we switch off the interactions and analyse the final state of the two spins via the state sensitive detection described above. In a sequence of experiments at constant $B_x$ we increase $T$ and therefore $J(t)/B_x$. After 50 steps of 2.5 $\micro$s each we reach the maximal amplitude $J(t=125\ \micro\text{s})/B_x =J_\text{max}/B_x=5.2$ (see methods) and achieve a quantum-magnetisation $\mathcal M=P_{\downarrow \downarrow}+P_{\uparrow \uparrow}$, the probability of being in a state with ferromagnetic order, of $\mathcal M=(98\pm2)\%$.

\begin{figure}
\includegraphics[width=9cm]{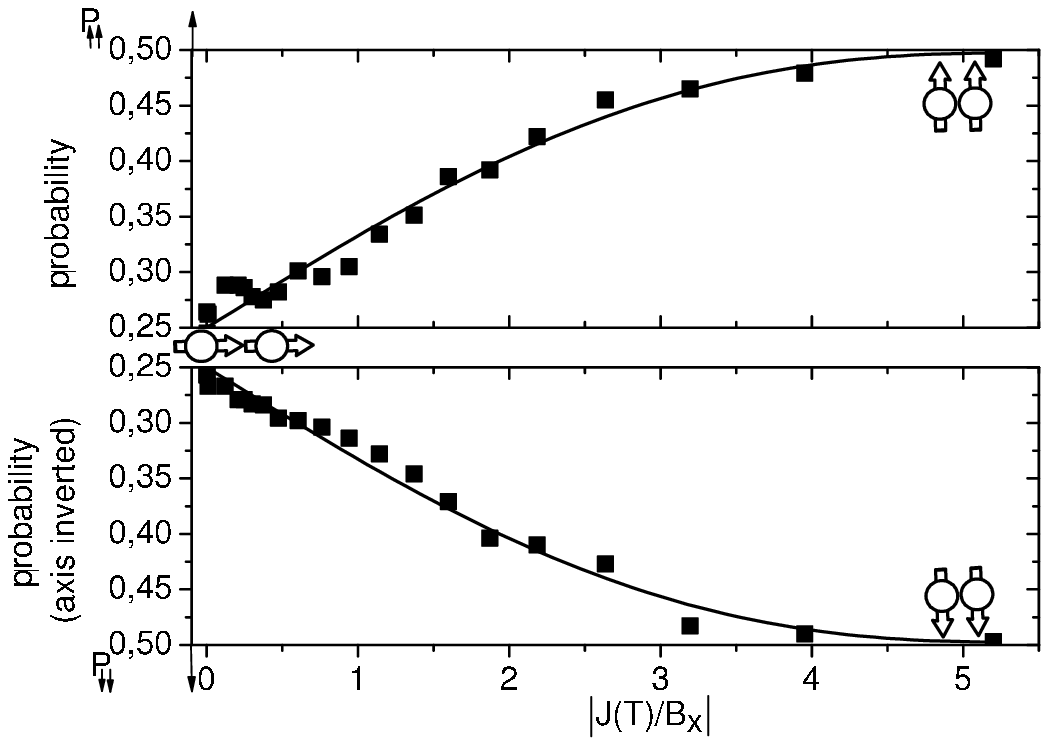}
\caption{\label{gabel} Quantum magnetisation of the spin system:
We initialise the spins in the paramagnetic state $\left| \rightarrow \right\rangle \left| \rightarrow \right\rangle = (\left| \uparrow \right\rangle+\left| \downarrow \right\rangle) (\left| \uparrow \right\rangle+\left| \downarrow \right\rangle) = \left| \uparrow \uparrow \right\rangle+\left| \uparrow \downarrow \right\rangle+\left| \downarrow \uparrow \right\rangle+\left| \downarrow \downarrow \right\rangle$, the ground state of the Hamiltonian $H_\text B=B_x  (\sigma_1^x+\sigma_2^x)$. A measurement of this superposition state would already project into each $\left| \uparrow \uparrow \right\rangle$ and $\left| \downarrow \downarrow \right\rangle$ with a probability of 0.25. 
After applying $B_x$ we adiabatically increase the effective spin-spin interaction $J(t=0)=0$ to $J(T)$. State sensitive fluorescence detection allows to distinguish the final states $\left| \uparrow \uparrow \right\rangle$, $\left| \downarrow \downarrow \right\rangle$, $\left| \uparrow \downarrow \right\rangle$ or $\left| \downarrow \uparrow \right\rangle$. Averaging over $10^4$ experiments provides us with its probability distribution $P_{\downarrow \downarrow}$ (two ions fluoresce), $P_{\uparrow \uparrow}$ (no ions fluoresces), and $P_{\uparrow \downarrow}$ or $P_{\downarrow \uparrow}$ (one ion fluoresces). We repeat the measurement for increasing ratios $J(T)/B_x$. The experimental results for the ferromagnetic contributions $P_{\uparrow \uparrow}$ and $P_{\downarrow \downarrow}$ are depicted as squares, the solid lines representing the theoretical prediction. For $J(T)/B_x \ll1$, the paramagnetic order is preserved. For $J(T)/B_x \gg1$, the spins undergo a transition into the ferromagnetic order, the ground state of the Hamiltonian $H_\text{J} =  J_\text{max} \sigma_1^z \sigma_2^z$, with a related quantum magnetisation $\mathcal M=P_{\downarrow \downarrow}+P_{\uparrow \uparrow}$ of $\ge (98\pm2)\%$. Note that we invert the ordinate of the lower frame to emphasise the unbroken symmetry of the evolution.}
\end{figure}

\section{Entanglement}
In our experiment we can detect both ferromagnetic contributions, $P_{\uparrow \uparrow}$ and $P_{\downarrow \downarrow}$, separately. Any imperfection in the simulation acting as a bias field $B_z$ along the $z$-axis, would energetically prefer one of the ferromagnetic states over the other and therefore unbalance their contribution to the final state. We carefully cancel all bias fields (see methods) to balance the populations $P_{\uparrow \uparrow}$ and $P_{\downarrow \downarrow}$, as can be seen in FIG.~\ref{gabel}. The results are in good agreement with theoretical predictions for our experiment, shown as solid lines. We expect the final state to be a coherent superposition of the two ferromagnetic states $\left|\uparrow \uparrow \right\rangle + \left|\downarrow \downarrow \right\rangle$, close to a maximally entangled Bell state. To quantify the experimentally reached coherence we measure the parity \cite{20} $\mathcal P= P_{\downarrow \downarrow} + P_{\uparrow \uparrow} -(P_{\downarrow \uparrow} +P_{\uparrow \downarrow})$ after applying an additional $R(\pi/2,\phi)$-pulse to both ions after $J_\text{max}$ is reached, with a variable rf-phase $\phi$ relative to the rf-field simulating $B_x$. 
\begin{figure}
\includegraphics[width=9cm]{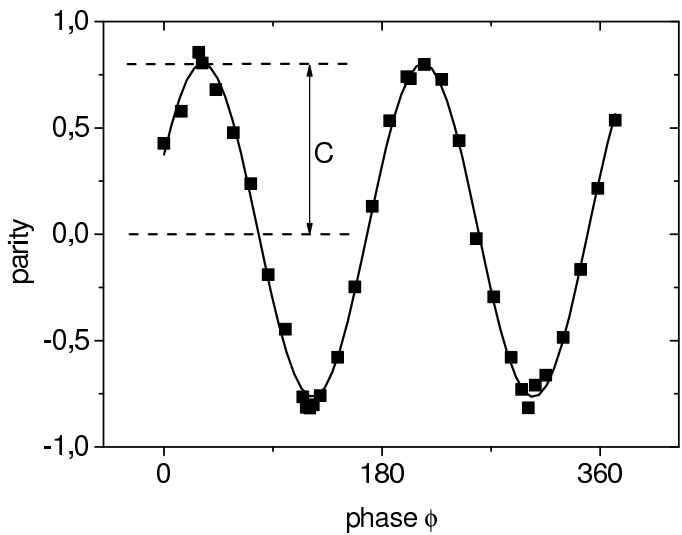}
\caption{\label{parflop} Entanglement of the quantum magnet: Measurement of the parity $\mathcal P= P_{\downarrow \downarrow} + P_{\uparrow \uparrow} -(P_{\downarrow \uparrow} +P_{\uparrow \downarrow})$ of the final ferromagnetic state after the simulation reached $J_\text{max} /B_x =5.2$. As we vary the phase $\phi$ of a subsequent analysis pulse, the parity of the two spins oscillates as $C \cos (2\phi)$.  Together with the final state populations $P_{\downarrow \downarrow}$ and $P_{\uparrow \uparrow}$ depicted in FIG.~\ref{gabel}, we can deduce a lower bound for the fidelity $\mathcal F= 1/2(P_{\downarrow \downarrow} + P_{\uparrow \uparrow}) + C/2$ of $(88\pm3)\%$ for the final superposition state $\left|\uparrow \uparrow \right\rangle +\left|\downarrow \downarrow \right\rangle$, a maximally entangled state, highlighting the quantum nature of this transition. We find qualitatively comparable results for the antiferromagnetic case $\left|\uparrow \downarrow \right\rangle + \left|\downarrow \uparrow \right\rangle$. Each data point averages $10^4$ experiments.}
\end{figure}
The measured data shown in FIG.~\ref{parflop} have a component that oscillates as $C \cos(2\phi)$, where $|C|/2$ characterises the coherences between the $\left|\uparrow \uparrow \right\rangle$ and $\left|\downarrow \downarrow \right\rangle$ components in the state produced. Deducing a contrast $C$ of $(78\pm2)\%$ from the best-fit we derive a lower bound of the fidelity \cite{20} $\mathcal F= 1/2(P_{\downarrow \downarrow} + P_{\uparrow \uparrow}) + C/2$ of $(88\pm3)\%$.

We also simulate the adiabatic evolution of a system not initialised in the ground state of the initial Hamiltonian. In particular we prepare the paramagnetic eigenstate $\left|\leftarrow \leftarrow \right\rangle= (\left|\downarrow \right\rangle -\left|\uparrow \right\rangle) (\left|\uparrow \right\rangle -\left|\downarrow \right\rangle)$, with the spins aligned anti-parallel with respect to the simulated magnetic field via a $R(\pi/2,\pi/2)$ rf-initialisation pulse. The adiabatic evolution should preserve the spin system in its excited state leading now into the anti-ferromagnetic order $\left|\uparrow \downarrow \right\rangle + \left|\downarrow \uparrow \right\rangle$. After evolution to $J=J_\text{max}$ we find $P_{\downarrow \uparrow} +P_{\uparrow \downarrow} \ge (95\pm2) \%$. To investigate the coherence between the $\left|\downarrow \uparrow \right\rangle$ and the $\left|\uparrow \downarrow \right\rangle$ components, we rotate the state via an additional $R(\pi/2,0)$-pulse which would ideally take $\left|\uparrow \downarrow \right\rangle + \left|\downarrow \uparrow \right\rangle \longrightarrow  \left|\uparrow \uparrow \right\rangle + \left|\downarrow \downarrow \right\rangle$, before we continue to measure the parity as explained above. We deduce a lower bound for the fidelity of the anti-ferromagnetic entangled state $\mathcal{F}= \left| \left\langle \Psi_\text{final} \big| \downarrow \uparrow + \uparrow \downarrow \right\rangle \right|^2=1/2 \left( P_{\uparrow \downarrow} + P_{\downarrow \uparrow} \right) +C/2$ of $(80\pm4)\%$. 

An equally valid viewpoint of this experiment interprets $\left|\leftarrow \leftarrow \right\rangle$ as the ground state of the Hamiltonian $-H_\text{Ising}$. Because the sign of all spin-spin interactions is also reversed in $-H_\text{Ising}$ it is equivalent to a change of sign in the spin-spin interaction $J$.

The entanglement of the final states additionally confirms that the transition from paramagnetic to (anti-)\linebreak{}ferromagnetic order is not caused by thermal fluctuations driving thermal phase transitions. The evolution is coherent and quantum mechanical, the coherent counterpart to the so-called quantum fluctuations \cite{3,21} driving quantum phase transitions in the thermodynamic limit. In this picture tunnelling processes \cite{21} induced by $B_x$ coherently couple the degenerate (in the rotating frame) states $\left| \uparrow \right\rangle$ and $\left|\downarrow \right\rangle$ with an amplitude proportional to ($B_x/|J|$). In a simplified picture for $N$ spins the amplitude for the tunnelling process between $\Psi_{N\uparrow}=\left| \uparrow \uparrow \ldots \uparrow \right\rangle$ and $\Psi_{N\downarrow}=\left| \downarrow \downarrow \ldots \downarrow \right\rangle$ is proportional to $(B_x/|J|)^N$, since all $N$ spins must be flipped. In the thermodynamic limit ($N \rightarrow \infty$) the system is predicted to undergo a quantum phase transition at $| J |  = B_x$. At values $J > B_x$ the tunnelling between $\Psi_{\infty \uparrow}$ and $\Psi_{\infty \downarrow}$ is completely suppressed. In our case of a finite system $\Psi_{2 \uparrow}$ and $\Psi_{2 \downarrow}$ remain coupled and the sharp quantum phase transition is smoothed into a gradual change from paramagnetic to (anti-)ferromagnetic order.

\section{Conclusion and Outlook}
We demonstrated the feasibility of simple quantum simulations in an ion trap by implementing the Hamiltonian of a quantum magnet undergoing a robust transition from a paramagnetic to an entangled ferromagnetic or anti-ferromagnetic order.  While our system is currently too small to solve classically intractable problems, it uses an approach that is complementary to a universal quantum computer in a way that can become advantageous as the approach is scaled to larger systems. Since our scheme only requires inducing the same overall spin-dependent optical force on all the ions \cite{2}, it does not rely on the use of sequences of quantum gates, thus its scaling to a higher number of ions can be simpler. Furthermore the desired outcome might not be affected by decoherence as drastically as typical quantum algorithms, because a continuous loss of quantum fidelity might not spoil completely the outcome of the experiment (for example the (anti-)ferromagnetic ordering transitions are hardly affected by phase decoherence), while universal quantum computation will almost certainly require involved sub-agorithms for error correction \cite{6}. Decoherence in the simulator might even mimic the influence of the natural environment \cite{23} of the studied system, if we judiciously construct our simulation (for example the decoherence we mainly observe in our demonstration implements a dephasing environment).

Despite technical challenges, we expect that this work is the start to extensive experimental research of complex many-body phases with trapped ion systems. Linear trapping setups may be used for the quantum simulation of quantum dynamics beyond the ground state where chains of 30 spins would already allow to outperform current simulations with classical computers. We may also adapt our scheme to new ion trapping technologies \cite{4}. For example, a modest scaling to systems of $20\times 20$ spins in 2D would yield insight into open problems in solid state physics, e.g. related to spin-frustration. This could pave the way to address a broad range of fundamental issues in condensed matter physics which are intractable with exact numerical methods, like, for example, spin liquids in triangular lattices, suspected to be closely related to phases of high-$T_\text c$ superconductors \cite{5}.

\section{Methods}
State dependent optical dipole force: An effective (Ising) spin-spin interaction was proposed to be implemented via magnetic field gradients \cite{24}. Porras and Cirac suggested to use state dependent optical dipole forces \cite{14,22} displacing the spin state $\left| s \right\rangle$ \mbox{($s$ either $\uparrow$ or $\downarrow$)} in phase space by an amount that depends on $\left| s \right\rangle$. The area swept in phase space changes the state to $\text{e}^{i \phi (s)} \left| s \right\rangle$. The phase $\phi(s)$ can be broken down into single spin terms proportional to $\sigma_i^z$ and apparent spin-spin interactions proportional to $\sigma_i^k \sigma_j^k$ and thus gives rise to the desired simulation of spin-spin interactions \cite{26}. It can also lead to single spin phases that simulate the unwanted contribution of a common bias fields $B_z \sigma^z$ in the Hamiltonian that will unbalance the probabilities $P_{\downarrow \downarrow}$ and $P_{\uparrow \uparrow}$. To achieve a balanced probability distribution as depicted in FIG.~3, we have to carefully compensate these single spin phases. To this end we (1) compensate the residual AC-Stark shifts of the individual laser beams by carefully choosing direction and polarisation of the beams \cite{13} and (2) compensate for the imbalance caused by single spin phases via a detuning of the order of several kHz of the rf-transition relative to $\omega_0$. (3) The ions have to be separated by an integer multiple of the effective wavelength $\lambda_\text{eff}= \lambda /\sqrt{2}$, in our implementation $18\times \lambda_\text{eff}$, requiring the control of the axial trapping frequency to better than 100 Hz. 
In contrast to phase gate implementations \cite{22} we have to detune the two laser beams far enough to keep the motional excitation and the related errors due to residual spin-motion coupling \cite{2} insignificant. Adjusting the detuning to $\delta=-(\omega_\text{stretch}-(\omega_1-\omega_2))=-2 \pi \times 250~\text{kHz}$ red of the stretch mode frequency and terminating the evolution after the system returned back into the motional ground state \cite{22} ideally completely cancels the simulation errors discussed in \cite{2}.

State sensitive detection: For two spins, the integrated fluorescence signal does not allow to distinguish between two ($\left| \uparrow \downarrow \right\rangle$ and $\left| \downarrow \uparrow \right\rangle$) of the four possible spin configurations. In addition, the amount of detected photons for each of the three distinguishable configurations fluctuates from experiment to experiment according to Poissonian statistics and therefore can only be determined with limited accuracy. For the data reported, we repeated each experiment $10^4$ times and fitted the resulting photon-number distribution to the weighted sum of three reference distributions to derive $P_{\downarrow \downarrow}, \ P_{\uparrow \uparrow}$ and $P_{\uparrow \downarrow} +P_{\downarrow \uparrow}$.

Adiabatic evolution: We achieve the best fidelities for the reported transitions at a duration of the simulation of $T=125\ \micro \text{s}$ at a $B_x = 2\pi \times 4.24 \ \text{kHz}$. Even though we are not strictly in the adiabatic limit, the robustness of the transition allows to minimise decoherence effects reducing the duration of the simulation. In addition, technical reasons led to the evolution of $J(t)$ linear in time to up to $J(t=50\ \micro \text{s}) = 5 \times 10^{-4} J_\text{max}$, continued by $J(t)\approx (\text e^{\alpha t}-\beta)^2$ best fitted by $\alpha=0.026$ and $\beta=4$. Up to now we did not improve the fidelities by evolving or terminating $J(t)$ or $B_x$ in a more adiabatic way.

\begin{acknowledgments}
This work was supported by the Emmy-Noether Programme of the German Research Foundation (DFG, Grant No. SCHA 973/1-2), the MPQ Garching, the DFG Cluster of Excellence Munich-Centre for Advanced Photonics, and the European project SCALA. We thank Dietrich Leibfried for his invaluable input and Ignacio Cirac and Gerhard Rempe for most interesting comments and generous support.
\end{acknowledgments}

\newpage
\bibliography{Magnet}

\end{document}